\documentstyle[preprint,prb,aps]{revtex}
\newcommand{\bmp}{{\mbox{\boldmath $p$}}}
\newcommand{\bms}{{\mbox{\boldmath $s$}}}

\newcommand{\bmr}{{\mbox{\boldmath $r$}}}

\newcommand{\bmq}{{\mbox{\boldmath $q$}}}

\begin{document}
\preprint {WIS-98/28 Oct-DPP; TRI-PP-98-29}
\draft

\date{\today}
\title{On the equivalence of the Impulse Approximation and the
Gersch-Rodriguez-Smith series for structure functions}
\author{A.S. Rinat$^{1,2}$ and B.K. Jennings$^2$}
\address{$^1$ Department of Particle Physics, Weizmann Institute of Science,
         Rehovot 76100, Israel\\
        $^2$ TRIUMF,Vancouver, B.C., Canada, V6T 2A3}

\maketitle
\begin{abstract}

We derive for a non-relativistic system an approximation for Final
State Interactions in a form, resembling a DWIA which corrects the
structure function computed in the PWIA.  We then compare the
Gersch-Rodriguez-Smith and the IA series for structure functions and
prove that to order ${\cal O}(1/q^2)$ the above DWIA representation is
contained in the GRS series to the same order.
There is an additional term in the GRS series that is
missing in the DWIA due to the eikonal approximations in the
latter. This strongly suggests that the two approaches, when
treated exactly, produce identical structure function to arbitrary
order in $1/q$.

\end{abstract}

\section{Introduction}

Virtually all computations of structure  functions of nuclei, as measured
by  inclusive  scattering  of  high-energy  electrons,  use  relativistic
generalizations of either the non-relativistic (NR), perturbative impulse
approximation (IA) series \cite{cio,go6}, or of a non-perturbative theory
\cite{rt1}, formulated by Gersch,  Rodriguez and Smith (GRS) \cite{grs1}.

We start  with the latter.  It  produces an expansion of  the response in
inverse powers of  the momentum transfer $q$  with coefficient functions,
depending on  many-body density-matrices which  are diagonal in  all, but
one coordinate.  Terms in that series  are the asymptotic limit for $q\to
\infty$  and  a  series  of  Final  State  Interaction  (FSI)  terms  for
decreasing, finite $q$.

The Impulse  Approximation (IA)  series is one  in the  interaction $\bar
V(\bmr_1)=\sum_{k\ge 2}V(\bmr_1-\bmr_k)$  between the struck  nucleon and
the core.  To lowest order, i.e.  in the Plane Wave Impulse Approximation
(PWIA),  one neglects  $\bar  V(\bmr_1)$.  FSI  interactions  for the  IA
series  are thus  perturbatively calculated  contributions of  increasing
order in the initially neglected interaction.

The  formal GRS  and  the IA  series appear  very  dissimilar, yet  those
provide two representations of the same structure function.  Consequently
an  $exact$ treatment  of each  ought  to produce  identical results.   A
frequently raised question is then, which approach is better when treated
$approximately$.  To our  knowledge not even a criterion,  to be followed
in principle, has  been formulated in the past.  The  main purpose of the
present  note  is  just  such  a formulation,  followed  by  a  proof  of
equivalence.

The above  quest is  encumbered by  the fact that  we do  not know  of an
exact, maneagable evaluation  of FSI in the IA series,  as exists for the
GRS theory.   A pre-requisite for  a comparison is therefore  a realistic
model for FSI, replacing the IA series.

As to  the nature  of such a  model one  is guided by  the fact  that the
relative weight  of FSI in  the response diminishes with  increasing $q$.
It  is  therefore natural  to  consider  kinematic conditions,  generally
reached for scattering  with high beam energies.  Those  are by necessity
accompanied by  effects due  to relativity,  particle production  and the
like, whose treatment can never be exact.  As a starting point we suggest
a well-defined  non-relativistic (NR) model,  based on a  hamiltonian for
point-particles, i.e. for  particles which cannot be  excited.  The model
can  be  treated  exactly  and   provides  insight  which  later  can  be
incorporated in realistic situations.

We start  in Section II with  the GRS theory, recapitulate  some formally
exact expressions for  the lowest order terms of the  GRS series and cite
results  for  partial summations  of  selected  higher order  terms.   In
Section  III   we  consider  the   response  of  a   semi-inclusive  (SI)
$A(e,e'p)X_{A-1}$  reaction  in the  PWIA,  which  features the  one-hole
spectral function.   We then suggest  a realistic  form for FSI  which in
nuclear  parlance is  called,  the Distorted  Wave Impulse  Approximation
(DWIA).  Integrating  the SI  response over the  momenta of  the outgoing
nucleon,  produces  for  that  model the  totally  inclusive  (TI)  cross
section.   We then  demonstrate  in Section  IV that  the  GRS to  ${\cal
O}(1/q^2)$ contains  the DWIA terms to  the same order and  attribute the
absence of  an extra term to  the approximate nature of  the chosen DWIA.
We conclude  that Section by comparing  ours with work of  similar scope.
In  Section  V  we briefly  discuss  the  embedding  of  the above  in  a
relativistic theory.

\section{The GRS series and some resummations.}

We consider the structure function for, or the response
$S(q,\omega)$ of, a NR many-body system to a scalar perturbation,
defined as
the ratio of the inclusive scattering by a projectile
and the elementary projectile-constituent cross section.
The kinematic variables $(q,\omega)$ are the momentum
and energy, transfered by the projectile to the target.
A form for the response per particle reads
\begin{eqnarray}
S(q,\omega)=(2\pi A)^{-1}\langle \Phi_A^0|\rho_q^{\dagger}
\delta(\omega-H_A)\rho_q|\Phi_A^0\rangle,
\label{a1}
\end{eqnarray}
where $H_A,\Phi_A^0,E_A^0$ are the exact $A$-body hamiltonian, its ground
state wave function and the corresponding energy.

For large $q$ it is useful to introduce the reduced response
$\phi(q,y)=(q/M)S(q,\omega)$, with $M$ the mass of a  particle and
$y$ a kinematic variable, replacing  the energy loss $\omega$
\cite{grs1,west}
\begin{eqnarray}
y=\frac {M}{q}\bigg (\omega-\frac {q^2}{2M}\bigg ),
\label{a2}
\end{eqnarray}
Substitution  of $\rho_q=\sum_j\,e^{i\bmq.\bmr_j}$  into  (1) produces
incoherent and coherent components. When considering
high-$q$ responses, it suffices to consider the dominant incoherent part,
where a single particle is tracked in its
propagation through the medium \cite{west}. We cite
Ref. \onlinecite{grs1} for a derivation of the GRS series
\begin{eqnarray}
\phi(q,y)=\sum_{n\ge 0} (1/v_q)^nF_n(y),
\label{a3}
\end{eqnarray}
where  $v_q=q/M$ is  the  recoil velocity,  corresponding  to a  momentum
transfer $q$.  The coefficient functions  $F_n(y)$ are functionals of the
inter-particle interaction $V$ and density matrices
$\rho_n(1'j;1j),\,j\ge  2$. Those are diagonal in all coordinates
$j=\bmr_j$, except that
of the  struck nucleon,  which is  chosen to be  $'1'$.  All  derive from
$\rho_A(1',k;1,k)\,,A\ge  k\ge  2$  and  satisfy in  our  convention  the
relations
\begin{eqnarray}
\rho_n(1',2...n;1,2...n)&=&\frac{1}{(A-n)!}
\bigg (\Pi_{j=n+1}^A\int dj\bigg ) \rho_A(1'j;1j)
\nonumber\\
\rho_A(1'k;1k) &=&A!\Phi^*_A(1',k)\Phi_A(1,k)
\label{a4}
\end{eqnarray}
The appearance  of exact many-body  densities shows that
from the  onset the theory
accounts  for correlations  of  the target  nucleons.

For our purposes it sufficies to mention $\rho_n$ for $n=1,2,3$
which enter expressions for the
asymptotic limit $F_0$ and the two dominant FSI corrections  $F_1,F_2$

\begin{mathletters}
\label{a5}
\begin{eqnarray}
F_0(y) &=&\frac {1}{A!}\int \frac{ds}{2\pi} e^{iys}\,\int d1
\bigg (\Pi_{k\ge 2}^A \int dk\bigg )
\rho_A(1-s,k;1,k)
\nonumber\\
&=&\frac {1}{A}\int \frac{ds}{2\pi}e^{isy} \int d1 \rho_1(1-s;1)
=\frac {1}{4\pi^2}\int_{|y|}^{\infty} dp\, p\, n(p)
\label{a5a}\\
\frac{1}{v_q}F_1(y)
&=& \frac{i}{A!} \int\frac {ds}{2\pi}e^{iys}
\int d1 \bigg \lbrack \Pi_{k\ge 2}^A \int dk \bigg \rbrack
\rho_A(1-s,k;1k)\sum_{k\ge 2} \tilde\chi_q(1-k,s)
\nonumber\\
&=& \frac{i}{A} \int\frac {ds}{2\pi}e^{iys}
\int \int d1\,d2
\rho_2(1-s,2;12)\tilde\chi_q(1-2,s)
\label{a5b}\\
\frac{1}{v_q^2}F_2(y)&=&-\frac {1}{2A!}\int\frac{ds}{2\pi}e^{isy}
\int d1 \bigg \lbrack \Pi_{k\ge 2}^A \int dk \bigg \rbrack
\rho_A(1-s,k;1,k)\bigg \lbrack\int_0^s d\sigma \sum_{k\ge 2}
\tilde\chi_q(1-k,s)\bigg\rbrack^2+\frac{1}{v_q^2}F_2^{(r)}(y)
\label{a5c}\\
\frac{1}{v_q^2}F_2^{(r)}(y)
&=&-\frac{1}{A!}\int\frac{ds}{2\pi}e^{isy}
\int d1 \bigg \lbrack \Pi_{k\ge 2}^A \int dk \bigg \rbrack
\rho_A(1-s,k;1,k)\bigg \lbrack \frac{1}{2}\frac{\partial^2}{\partial s^2}
\bigg (\sum_{k\ge 2} \int_0^s d\sigma \tilde\chi_q(1-k,\sigma)\bigg )^2
\nonumber\\
&&-\bigg (\sum_{k\ge 2}\tilde\chi_q(1-k,s)\bigg )^2 \bigg\rbrack
\label{a5d}
\end{eqnarray}
\end{mathletters}
with $n(p)$ the single-particle
momentum distribution. The expression for $F_2^{(r)}$
is easily derived from Eq. (14) in Ref. \onlinecite{rt3}.
Above we also introduced
\begin{eqnarray}
\tilde\chi_q(1,s)&=& \tilde\chi_q^{(1)}(1,s)+\tilde\chi_q^{(2)}(1,s)
\nonumber\\
&=&-(1/v_q) \int _0^s d\sigma [V(1-\sigma)-V(1)],
\label{a6}
\end{eqnarray}
where we write symbolically
\begin{eqnarray}
{\cal V}&=&{\cal V}^{(1)}+{\cal V}^{(2)}
\nonumber\\
{\cal V}^{(1)} \to {\cal V}^{(1)}_{\sigma}& = & V(1-\sigma)
\nonumber\\
{\cal V}^{(2)}&\to& -V(1)
\label{a7}
\end{eqnarray}
Eq. (\ref{a6}) defines the  coordinate representation of
the off-shell eikonal  phase $\tilde\chi(1,s)$
corresponding to the total ${\cal V}$ and
its components $\tilde\chi^{(a)}(1,s)\,;a=1,2$ which
are characteristic of  the
GRS theory or of path integral  methods \cite{car}.

It is frequently useful to make resummations
within the GRS series (\ref{a3}).
We consider first a ladder sum of  repeated interactions
$V$ which results in the replacement $V \to  t=V_{eff}$.
This replacement is mandatory if the bare interaction $V$
is singular.
The  corresponding  change in the phase $\tilde\chi$ (\ref{a6}) is
\begin{eqnarray}
i\tilde \chi \to \tilde \Gamma\equiv e^{i\tilde\chi}-1,
\label{a8}
\end{eqnarray}
with $\tilde \Gamma$, the total off-shell profile function.

Next we consider a
cumulant resummation which to lowest order reads \cite{grs2}
\begin{eqnarray}
\phi(q,y)=\frac{1}{A}\int\frac{ds}{2\pi}e^{isy}\int d1 \rho_1(1-s;1)
{\rm exp}\bigg \lbrack\frac {\int d2\rho_2(1-s,2;1,2)
\tilde\Gamma_q(1-2,s)}
{\rho_1(1-s;1)}+....\bigg \rbrack
\label{a9}
\end{eqnarray}
When expanded, it reproduces
the lowest order terms in the GRS series, as well as selected higher
order contributions (\ref{a3}).

\section{FSI corrections to the PWIA response.}

By way of introduction we consider first SI scattering.
The corresponding response per nucleon is
\begin{eqnarray}
S^{SI}(q,\omega;\bmp)=\frac{1}{A}\sum_m \bigg |\bigg \langle
\Phi_A^0|\rho_q^{\dagger}|\Psi_{(A-1)_m;\bmp+\bmq}^{(-)}
\bigg \rangle \bigg |^2\delta\bigg (\omega-\Delta_m-
\frac {(\bmp+\bmq)^2}{2M}\bigg ),
\label{a10}
\end{eqnarray}
where $\bmp$ is the momentum of the struck, and
$\bmp+\bmq$ that of the
detected outgoing nucleon after absorbing the momentum transfer $\bmq$.
$\Psi^{(-)}_{{(A-1)_m};\bmp+\bmq}$  is the  state
of that nucleon, scattered from a nucleus of $A-1$ particles in
state $m$. $\Delta_m$ is the separation energy of a nucleon in the
ground state $A$-body system, with the  daughter nucleus
in the state $\Phi_{A-1}^m$. We write the total hamiltonian as
\begin{eqnarray}
H_A(1;k)=H_{A-1}(k)+T(1)+{\bar V}(1)
\label{a11}
\end{eqnarray}
with ${\bar V}(1)=\sum_{k\ge 2}V(1-k)$, the interaction
of particle $'$1$'$ with the core. Neglect of the latter defines the PWIA
\begin{eqnarray}
\bigg \lbrack \Psi^{(-)}_{{(A-1)_m};\bmp+\bmq}(\bmr_1;\bmr_k)
\bigg\rbrack ^{PWIA}
\to \Phi_{A-1}^m(\bmr_k) e^{-i(\bmp+\bmq).\bmr_1}
\label{a12}
\end{eqnarray}
When substituted into Eq. ({\ref{a10}), it produces the standard PWIA
approximation for the SI response
\begin{mathletters}
\label{a13}
\begin{eqnarray}
S^{SI;PWIA}(q,\omega;\bmp)&=&\int dE P(\bmp,E)
\delta\bigg (E-\omega-\frac{(\bmp+\bmq)^2}{2M}\bigg )
\label{a13a}\\
\phi^{SI;PWIA}(q,y_0;\bmp)&\approx &\delta(y_0-p_z)n(p)
\label{a13b}\\
n(p)&=&\int dE P(\bmp,E)
\label{a13c}
\end{eqnarray}
\end{mathletters}
Here $P(\bmp,E)$ is  the single-hole spectral function,  dependent on the
separation-energies of each of the daughter states $m$.  Eq. (\ref{a13b})
results from  the  assumption  that those  separation
energies  may  be
replaced by an average $\Delta_m\to \langle \Delta\rangle$.  One may then
replace the  energy loss  $\omega$ by  the IA  scaling variable,  also in
terms of $\langle\Delta\rangle$
\begin{eqnarray}
y_0=-q+\sqrt{2M(\omega-\langle \Delta \rangle)}
\label{a14}
\end{eqnarray}
FSI  corrections  to the  PWBA  result  (\ref{a13b}) are  by  definition,
contributions   due   to   the    residual   interaction   $V$,   treated
perturbatively.  With no practical way
to do so systematically, we proceed in
an approximative  manner and  assume that  the outgoing  nucleon scatters
from an initially frozen configuration of $k$ nucleons \cite{ja3,rj,rt2}
\begin{eqnarray}
\Psi_{{(A-1)_m};\bmp+\bmq}^{(-)}(1;k)\approx
\Phi_{A-1}^m(k)\psi^{(-)}_{\bmp+\bmq}(1;\langle k \rangle),
\label{a15}
\end{eqnarray}
Such an approximation is justified if, with
respect to the Fermi momentum $p_F$
\begin{eqnarray}
p&\approx& p_F
\nonumber\\
|\bmp+\bmq|&\approx& q\gg p_F,
\nonumber
\end{eqnarray}
One notes that, contrary to the
perturbative nature of the actual IA series, the
approximation  (\ref{a15}) for it is non-perturbative.

Substituting  (\ref{a15}) into (\ref{a10}) and replacing again
state-dependent separation energies by an average, one performs
closure over states of the daughter nucleus and obtains
\begin{eqnarray}
\phi^{SI}(q,y_0;\bmp)&\approx&  \delta(y_0-p_z)\bigg\langle
\Phi_A^0(1';k)|e^{-i\bmq.\bmr_1'}|\psi^{(-)}_{\bmp+\bmq}(1';k)
\bigg \rangle \bigg \langle \psi^{(-)}_{\bmp+\bmq}(1;k)|e^{i\bmq.\bmr_1}|
\Phi_A^0(1;k)\bigg \rangle ^*
\nonumber\\
&\approx&
\delta(y_0-p_z)\bigg\langle \Phi_A^0(1';k)|
e^{-i\bmp.\bmr_1'}|\xi^{(-)}_{\bmp+\bmq}(1';\langle k \rangle)
\bigg \rangle
\bigg \langle \xi^{(-)}_{\bmp+\bmq}(1;\langle k\rangle|
e^{i\bmp.\bmr_1}|
\Phi_A^0(1;k)\bigg \rangle^*
\label{a16}
\end{eqnarray}
In Eq. (\ref{a16})  we used the standard eikonal expression
for the state describing scattering from a static field with fixed
coordinates $\langle k\rangle$
\begin{eqnarray}
\psi^{\pm}_{\kappa}(1; \langle k \rangle)=
e^{i\kappa z_1}\xi^{\pm}_{\kappa}(1;\langle k\rangle)
\label{a17}
\end{eqnarray}
due to the static non-central field $\sum_{k\ge 2}V(1-\langle k\rangle)$.
The distortion function $\xi$ in the approximation $|\bmp+\bmq|
\approx q$   reads \cite{gl}
\begin{eqnarray}
\xi_q^{(-)}(1;\langle k\rangle)={\rm exp}\bigg\lbrack-\frac{i}{v_q}\sum_k
\int_{z_1}^{\infty} d\zeta
V(1-\langle k\rangle-\zeta)\bigg \rbrack
\label{a18}
\end{eqnarray}
After substitution into (\ref{a16}) one restores the core coordinates to
their dynamical status and obtains for real $V$
the following expression for the SI response in the DWIA
\begin{eqnarray}
\phi^{SI,DWIA}(q,y_0;\bmp)&=&\frac{1}{A!}\delta(y_0-p_z)
\int d\bms e^{i\bmp.\bms}\int d1 \Pi_{k \ge 2} dk \rho_A(1-s,k;1,k)
\nonumber\\
&&{\rm exp}\bigg \lbrack -\frac {i}{v_q}\sum_k\int_{z_1'}^{z_1}
d\zeta V(1-k-\zeta)\bigg \rbrack
\label{a19}
\end{eqnarray}
At this stage one exploits  the  absence of degrees
of freedom, others than point-particles.
The TI response is then obtained  by integrating
the SI response  over  missing momenta $\bmp$.
As a result $\bms=\bmr_1-\bmr_1'= s \hat \bmq$ lies in the
direction of $\bmq$ and one finds
\begin{eqnarray}
\phi^{TI,DWIA}(q,y_0)= \frac{1}{A!}
\int\frac{ds}{2\pi}e^{iy_0s}  \int d1 \Pi_{k \ge 2}\int dk
\rho_A(1-s,k;1,k) {\rm exp}[i\sum_{k\ge 2}\tilde\chi^{(1)}_q(1-k,s)]
\label{a20}
\end{eqnarray}
The above result has still the full complexity of a many-body problem,
present in the
$A$-body density matrix. That complexity  is considerably reduced
in  an independent-pair (Kirkwood) approximation
\begin{eqnarray}
\rho_A(1k;1'k)\approx \frac {(A-1)!}{(A-1)^{A-1}}
\frac{\Pi_{k\ge 2}^A\rho_2(1k;1'k)}{[\rho_1(1;1')]^{A-2}},
\label{a21}
\end{eqnarray}
which respects the sum rules (\ref{a4}).
Substitution in (\ref{a20})
produces for the reduced TI response per nucleon in the DWIA
\begin{mathletters}
\label{a22}
\begin{eqnarray}
\phi^{TI,DWIA}(q,y_0)&\approx&
\frac {1}{A(A-1)^{A-1}}\int \frac{ds}{2\pi}e^{iy_0s}
\frac {\int d1}{[\rho_1(1-s;1)]^{A-2}}
\nonumber\\
&&\bigg \lbrack \Pi_{k \ge 2}\int  dk
\rho_A(1-s,k;1,k) \tilde\Gamma^{(1)}_q(1-k,s)\bigg \rbrack
\nonumber\\
&\approx &\frac {1}{A}\int \int \frac{ds}{2\pi}e^{iy_0s} d1\rho_1(1-s;1)
{\rm exp}\bigg \lbrack
\frac{\int\,d2 \rho_2(1-s,2;1,2)\tilde \Gamma_q^{(1)}(1-2,s)}
{\rho_1(1-s;1)}\bigg \rbrack		
\label{a22a}\\
&\approx& \frac{1}{A} \int dy_0'\, F_0(y_0-y_0'){\cal R}_q(y_0')
\label{a22b}\\
{\cal R}_q(y_0)&\approx&
\int \frac{ds}{2\pi}e^{iy_0s} \int d1
{\rm exp}\bigg \lbrack
\frac{\int\,d2 \rho_2(1-s,2;1,2)\tilde \Gamma_q^{(1)}(1-2,s)}
{\rho_1(1-s;1)}\bigg \rbrack
\label{a22c}
\end{eqnarray}
\end{mathletters}
For later use we expressed the response (\ref{a22a}) as a convolution
of the asymptotic limit and a generalized FSI factor (cf. Eq. (5a), of
the last Ref. \onlinecite{rt1}).

There clearly is a formal similarity in the expressions
(\ref{a9}) and (\ref{a22a}) for the TI response in,
respectively,  the first
cumulant expression in the GRS theory, and the approximate IA series.
The apparent differences amount to i) the appearance of $y_0$ instead of
$y=y_w$ and ii) the presence in the DWIA of a profile function
$\tilde\Gamma^{(1)}$, related  to the first potential in
(\ref{a6}) and not to both, as
is the case in the GRS theory.  In the following Section we shall
investigate whether, and to what extend these
apparently similar expressions coincide.

\section{Measure of equivalence of GRS and approximate IA series.}

There are two, in principle equivalent
ways to compare the $exact$ IA and GRS series for the response,
namely by isolating and counting
powers in either the residual interaction $\bar V(1)$
or in $1/q$. However, in view of the fact that the IA series is treated
approximately, the exact GRS series becomes the
natural standard. Both
approaches shall be traced to terms up, and including ${\cal O}(1/q^2)$.

We start with the GRS series (\ref{a3})
\begin{eqnarray}
\phi(q,y;{\cal V})=\sum_{n\ge 0}(1/v_q)^nF_n(y;{\cal V})
\label{a23}
\end{eqnarray}
Using (\ref{a7}) we make explicit the two components of
the $'$total$'$ interaction action  in (\ref{a5})
\begin{mathletters}
\label{a24}
\begin{eqnarray}
F_0(y)&=&\frac {1}{4\pi^2}\int_{|y|}^{\infty} dp\, p\, n(p)
\label{a24a}\\
F_1(y)
&=&-\frac{i}{A!}\int\frac {ds}{2\pi}e^{iys}
\int d1 \bigg \lbrack \Pi_{k\ge 2}^A \int dk \bigg \rbrack
\rho_A(1-s,k;1k)
\nonumber\\
&&\sum_{k\ge 2}\int_0^s d\sigma
[V(1-2-\sigma)-V(1-k)]
\nonumber\\
&=& -\frac{1}{A} \int\frac {ds}{2\pi}e^{iys}
\int \int d1\,d2 \rho_2(1-s,2;12)\int_0^s d\sigma
[V(1-2-\sigma)-V(1-2)]
\label{a24b}\\
F_2(y)&=& \frac {i^2}{A!}\int\frac{ds}{2\pi}e^{isy}
\int d1 \bigg \lbrack \Pi_{k\ge 2}^A \int dk \bigg \rbrack
\rho_A(1-s,k;1k)
\nonumber\\
&&\frac{1}{2}\bigg \lbrack\int_0^s d\sigma \sum_{k\ge 2}
\int_0^s d\sigma[V(1-k-\sigma)-V(1-k)]\bigg\rbrack^2+F_2^{(r)}(y)
\label{a24c}
\end{eqnarray}
\end{mathletters}
Next we get to  the approximate DWIA expression (\ref{a22a}), which is of
non-perturbative nature but, which using ({\ref{a6})  may be formally
expanded
\begin{eqnarray}
\phi^{TI,DWIA}&&(q,y_0;{\cal V}^{(1)})=\frac{1}{A}
\int \frac{ds}{2\pi}e^{iy_0s}
\int d1\bigg \lbrack \rho_1(1-s;1)-
\nonumber\\
&&\frac{i}{v_q}\int d2\rho_2(1-s,2;1,2)
\int_0^s d\sigma V(1-2-\sigma)
-\frac{1}{2v_q^2}\int d2 \bigg (\int_0^s d\sigma V^{(1)}(1-2-\sigma)
\bigg )^2-
\nonumber\\
&&\frac{1}{2v_q^2}\int d2\int d3
\rho_3(1-s,2,3;1,2,3) \int_0^s d\sigma
V(1-2-\sigma)\int_0^s d\sigma' V(1-3-\sigma')
\bigg \rbrack +{\cal O}(1/v_q^3)
\label{a25}
\end{eqnarray}

It is then our quest to investigate  whether, and to what extent,
the terms (\ref{a24a}) - (\ref{a24c})
of the GRS series contain the DWBI counterparts (\ref{a25})
. We do so by the following technique:

i) Separate in Eqs. (\ref{a24}) terms which depend exclusively
on ${\cal V}^{(1)}$. The former we expect to meet in (\ref{a25}).

ii) Track in the remainder of (\ref{a24}) parts where ${\cal V}^{(2)}$
acts on the groundstate in $\rho_A$. Using (\ref{a9})  one has
\begin{eqnarray}
\bigg \lbrack\sum_{l\ge 2}V(1-k)\bigg \rbrack\Phi(1,k)&=&
[H_A-H_{A-1}-T(1)]\Phi(1,k)
\nonumber\\
&\approx&-\int\frac{d\bmp}{2\pi^3}e^{i\bmp.\bmr}\bigg
(\langle \Delta\rangle +\frac{p^2}{2M}\bigg )\Phi(\bmp,k),
\label{a26}
\end{eqnarray}
where, in line with assumption made above,
separation energies are again replaced by an average.

iii) Collect terms, which enable the
replacement of the GRS-West scaling variable by the IA one, making use of
\begin{eqnarray}
y(=y_w)=y_0+\frac {1}{v_q}\bigg (\frac {y_0^2}{2M}+
\langle \Delta\rangle\bigg ),
\label{a27}
\end{eqnarray}

We start with
\begin{eqnarray}
F_1({\cal V})=F_1^{(1)}({\cal V}^{(1)})+F_1^{(2)}({\cal V}^{(2)}),
\label{a28}
\end{eqnarray}
where the superscripts indicate dependence on
${\cal V}^{(1)}, {\cal V}^{(2)}$ and following i) we consider the part
\begin{eqnarray}
\frac{1}{v_q}F_1^{(2)}(y)&=&-\frac{i}{A!}
\frac{\partial}{\partial y}\int \frac{ds}{2\pi}
e^{iys} \int d1[\Pi_{k\ge 2}\int dk]\rho_A(1-s,k;1,k)\frac{1}{v_q}
\sum_{l\ge 2}V(l)
\nonumber\\
&=&\frac{\partial}{\partial y}\int\frac {d\bmp}{2\pi^3}\delta(p_z-y)
\frac{1}{v_q}\bigg (\langle\Delta\rangle+\frac{p^2}{2M}\bigg )
n(p)
=(y-y_0)\frac{dF_0(y)}{dy}
\label{a29}
\end{eqnarray}

Continuing with $F_2$ we write (cf. Eqs. (\ref{a5c}), (\ref{a24c}))
\begin{eqnarray}
F_2({\cal V})
=F_2^{(1)}+F_2^{(2)}+F_2^{(1,2)}+F_2^{(r)},
\label{a30}
\end{eqnarray}
with $F_2^{(1,2)}$ containing mixed ${\cal V}^{(1)},
{\cal V}^{(2)}$ terms. The reasoning which leads to (\ref{a29})
produces
\begin{eqnarray}
\frac{1}{v_q^2}F_2^{(2)}(y)&=&\frac{1}{2A!}
\frac{\partial^2}{\partial y^2}\int \frac{ds}{2\pi}e^{iys}
\int d1[\Pi_{l\ge 2}\int dl]\rho_A(1-s,k;1,k)\frac{1}{v_q}
\sum_{k\ge 2}V(1-k) \frac{1}{v_q}\sum_{l\ge 2}V(1-l)
\nonumber\\
&=&\frac{1}{2}\bigg \lbrack\frac{1}{v_q}\bigg(\langle\Delta\rangle+
\frac{y^2}{2M}\bigg )^2\bigg \rbrack \frac{d^2 F_0(y)}{dy^2}
=\frac{1}{2}(y-y_0)^2\frac {d^2F_0(y)}{dy^2}
\label{a31}
\end{eqnarray}
Finally for the mixed term
\begin{eqnarray}
\frac{1}{v_q}F_2^{(1,2)}(y)&=&-\frac{1}{A!}\frac{\partial}{\partial y}
\int \frac{ds}{2\pi}e^{iys}
\int d1[\Pi_{l\ge 2}\int dl]\rho_A(1-s,k;1,k)
\frac{1}{v_q} \sum_{k\ge 2}V(1-k)
\nonumber\\
&&\frac{1}{v_q}\sum_{l\ge 2}\int_0^sd\sigma V(1-l-\sigma)]
\nonumber\\
&=&\frac{1}{v_q}\bigg (\frac {y^2}{2M}+
\langle\Delta\rangle\bigg )\frac {dF_1^{(1)}(y)}{dy}
=(y-y_0)\frac{dF_1^{(1)}(y)}{dy}
\label{a32}
\end{eqnarray}

Assembling the last three results and using (\ref{a27}) one finds
\begin{mathletters}
\label{a33}
\begin{eqnarray}
\phi(q,y;{\cal V})&=&
F_0(y)+\frac{1}{v_q}F_1(y;{\cal V})+\frac{1}{v_q^2}F_2(y;{\cal V})+
{\cal O}(1/v_q^3)
\label{a33a}\\
&=&\bigg \lbrack F_0(y)+ (y-y_0)\frac{dF_0(y)}{dy}+
\frac{1}{2}(y-y_0)^2\frac{{d^2}F_0(y)}{dy^2}\bigg\rbrack
\nonumber\\
&+&\frac{1}{v_q}\bigg \lbrack F_1^{(1)}(y)+(y-y_0)
\frac{dF_1^{(1)}(y)}{dy}\bigg\rbrack
+\frac{1}{v_q^2}\bigg \lbrack F_2^{(1)}(y)\bigg\rbrack +
\frac{1}{v_q^2}F_2^{(r)}(y)+ {\cal O}(1/v_q^3)
\label{a33b}\\
&=&\bigg\lbrack F_0(y_0)+\frac{1}{v_q}F_1^{(1)}(y_0)
+\frac{1}{v_q^2} F_2^{(1)}(y_0)\bigg\rbrack
+\frac{1}{v_q^2}F_2^{(r)}(y) +{\cal O}(1/v_q^3)
\label{a33c}\\
&=&\phi(q,y_0;{\cal V}^{(1)})+\frac{1}{v_q^2}F_2^{(r)}(y;{\cal V})
+{\cal O}(1/v_q^3)
\label{a33d}
\end{eqnarray}
\end{mathletters}
In Eqs. ({\ref{a33a}), (\ref{a33d}) we reinstated for greater clarity the
dependence on ${\cal V}$ and its component
${\cal  V}^{(1)}$.  The  above
demonstrates that all  terms of ${\cal O}(1/v_q^2)$ of the  IA series are
contained in  the GRS series  of the same  order, which however,  has one
additional term, not  reproduced in the DWIA.  This can  be traced to the
approximation (\ref{a15}).  Higher order  eikonal terms \cite{wal} to the
distortion function $\xi$ in (\ref{a18})  are at least of order $1/v_q^2$
and are expected to account for the above difference to that order.

Eqs.  (\ref{a29}), (\ref{a31})  and (\ref{a32})  are truly  remarkable in
that the  ${\cal V}^{(2)}$ dependence of  coefficient functions of  given
order can be expressed in  derivatives of functions $F$ of lower  order
$F$, and  which are  free of that interaction component. Grouped
terms ultimately produce the  replacement $y\to y_0$ and bring about
significant cancelations in the GRS series.

The above  completes the equivalence proof of the two  expressions of
the structure function for any NR  many-body system.  Were it not for the
use  of average  separation  energies, Eq.  (\ref{a33d})  would be  exact
\cite{rd}.  In a way this  approximation is unavoidable, because the
appearance of an essentially kinematic  IA scaling variable $y_0$  as in
Eq.  (\ref{a18})  requires  an  average separation  energy.   An  earlier
attempt   to  keep actual  separation   energies  invites   other
approximations (cf.   Eqs. (2.23) e.v.  in Ref. \onlinecite{rd}),
but we shall  not  pursue that  extension here.

We conclude this section, by mentioning previous incorporations of FSI
interactions for the IA series. In particular
Benhar and coworkers  \cite{om1,om2} advocated a convolution  of the PWIA
spectral function and some FSI,  specifically in the energy loss variable
$\omega$.   Their  FSI features  the  component  ${\cal V}^{(1)}$  as  in
({\ref{a22a}), based on our DWIA, and not the full ${\cal V}$.

We emphasize that a convolution of  the lowest order asymptotic limit and
a FSI factor has been $proved$ for the GRS theory and its series, and the
appropriate variable is  the GRS-West scaling variable  $y$ and not
$\,\,\omega$ \cite{grs2}. A presumed generalization, valid for the IA
series certainly requires a proof, which to our knowledge has not been
provided. Such a proof would
select the convolution variable.

Let us put aside the ad hoc convolution and attempt to replace
$\omega$ by a scaling variable. That is possible for any
candidate, built from purely kinematic variables as is $y_w$,
Eq. (\ref{a2}). The result is clearly neither (\ref{a9}) nor
(\ref{a22a}): The latter manifestly requires the IA variable (\ref{a27}),
but should be ruled out, because it assumes the existence
of an average separation energy, which counters the emphasis on the
exact spectral function with its state-dependent $\Delta_m$.
The latter is just demonstrated by
(\ref{a22a}),  which   may  be  written  as  a  convolution
(\ref{a22b}) in the $'$natural$'$ IA variable $y_0$.

Finally we mention  a, as yet unpublished conference report
\cite{pet} which also uses the above folding procedure. A discussion
should await its publication.

\section{Summary and conclusion.}

Our major goal above  was a comparison of the structure  function of a NR
system  of point-particles,  when computed  by means  of the  GRS and  IA
series.  Whereas for the former there exists a formally exact expression,
we do not know  of a similar, manageable
one for the IA series  beyond the PWIA.  Any
comparison therefore requires first  an approximation for FSI corrections
to the PWIA.  Having described a representative DWIA for the IA response,
we could perform the above comparison.

Our demonstration starts with of the GRS series up to and including
${\cal O}(1/v_q^2)$,  with  coefficients,  functions   of  the
GRS-West  scaling
variable $y$.  We  then proved striking cancelations,  producing the same
lowest order  terms from the  DWIA expressed  in the parallel  IA scaling
variable  $y_0$.  One  unretrieved  term in  the GRS  series is
undoubtedly due  to the chosen DWIA, approximating the FSI in the
actual IA series. Similar cancelations are expected to occur to any
order.

The above  success naturally  elicits the  question of  a relativistic
extension.  There  clearly is no  hope to derive results  with comparable
rigor.  It is  nevertheless of interest to recall here  some models where
nuclear  and nucleon  structure functions  are related  by a
generalized convolution \cite{gr}
$$ F^A=f^{PN}*F_N,$$
with $f^{PN}\propto \phi$. Here $\phi$ is in principle the structure
function of  a nucleus,  composed of point-particles,  where
inter-nucleon  potentials as in (\ref{a8}) are  replaced  by
scattering amplitudes
which have also meaning in a relativistic theory.
We    refer   to   Ref.
\onlinecite{rt1} where a generalization
of  the  effectively  2-component  interaction in  the  above  spirit  is
discussed.

It is  our hope  that this paper will lay  to rest  a long-lingering,
and occasionally controversial, issue in the study of responses.

\section{Acknowledgments.}

ASR is grateful to E. Mavrommatis for discussions on the topic and for
attending him to the work of Patraki et al. He also acknowledges the graceful
hospitality, experienced at TRIUMF.  BKJ thanks the Natural Sciences
and Research Council of Canada for financial support.

\newpage

\end{document}